\documentclass[pdflatex,sn-mathphys-num]{sn-jnl}
\pdfoutput=1

\usepackage{amsmath}
\usepackage{caption}
\usepackage{enumitem}
\usepackage{graphicx}
\usepackage{gensymb}
\usepackage{wrapfig}
\usepackage{color}
\usepackage{scrextend}
\usepackage{physics}
\usepackage{mathrsfs}
\usepackage{float}
\usepackage{bbm}
\usepackage{subcaption}
\usepackage{setspace}
\usepackage{lineno}

\begin{document}
	
\title{Asymmetrical cavity design that bypasses mode mixings in axion haloscope experiments}
\author[1]{\fnm{Byeongrok} \sur{Ko}}\email{brko@korea.ac.kr}
\author*[2]{\fnm{Andrew K.} \sur{Yi}}\email{andrewyi@slac.stanford.edu}

\affil[1]{\orgdiv{Department of Accelerator Science}, \orgname{ Korea University Sejong Campus}, \orgaddress{\city{Sejong}, \postcode{30019}, \country{Republic of Korea}}}
\affil*[2]{\orgname{SLAC National Accelerator Laboratory}, \orgaddress{\street{2575 Sand Hill Rd.}, \city{Menlo Park}, \postcode{94025}, \state{California}, \country{USA}}}

\abstract{Microwave cavities used in axion haloscope experiments typically employ a tuning rod as a means to widen the range of resonance frequencies at which it is sensitive to axion-to-photon conversion. A realistic tuning mechanism requires a gap between the cavity end caps and the tuning rod to ensure movement, and causes some modes to hybridize with the resonant mode that is being tracked for the experiment. These so-called mode mixings lead to gaps in the frequency range that practically lose sensitivity to axions. In order to solve this problem, we present a cavity design which, for two tuning rod configurations corresponding to a lower and higher frequency range, have a dielectric rod inserted at a specific location that makes the cavity asymmetrical. Moving the tuning rod closer to the dielectric insert changes the location and frequency of the mode mixing compared to when it is farther away from it. This design is easily realizable in practical experiments and makes possible an axion dark matter search with minimal loss in sensitivity due to mode mixings. We also show that the same design has the same desired effect when cavity dimensions are scaled down to be smaller and are at higher resonance frequencies.}

	
\maketitle

\section{Introduction}
	Axions \cite{Axion1, Axion2} are the resulting particles from a spontaneously broken symmetry introduced as a compelling solution for the strong $CP$ problem \cite{StrongCP1, StrongCP2, StrongCP3, StrongCP4, StrongCP5}, first formulated by Peccei and Quinn \cite{PQ}. Coincidentally, these particles are a candidate \cite{AxionCDM1, AxionCDM2, AxionCDM3} for cold dark matter \cite{CDM1, CDM2}, whose properties are unknown but which constitutes over 85\% of the matter in the universe \cite{Planck}. As this particle can potentially solve separate problems in both particle physics and cosmology, there is a strong motivation to create an experiment that can detect axions. Haloscopes, first developed by Sikivie \cite{Sikivie}, utilize the axion-to-photon conversion under a strong magnetic field. In order to increase the conversion power of the axion, many experiments use a cylindrical cavity which significantly enhances the conversion power of an axion signal around the resonance frequency of a cavity mode. Due to the unknown mass of the axion, a tuning rod is used to change the resonance frequency as its moves inside the cavity \cite{Cavity}, allowing access to a wider range of frequencies. A haloscope experiment is designed to scan through frequencies at a set sensitivity, such as those for benchmark axion models Kim-Shifman-Vainshtein-Zakharov (KSVZ) \cite{KSVZ1, KSVZ2} and Dine-Fischler-Srednicki-Zhitnitskii (DFSZ) \cite{DFSZ1, DFSZ2}, as quickly as possible. The figure of merit for such resonant experiments is the scanning rate $R$ \cite{ScanRate}. $R$ depends on various paremeters, but the relevant parameters for this work will be the quality factor $Q$ and form factor of the cavity mode, which is discussed below. As this work does not cover the effects of antenna coupling, we will use the cavity's unloaded quality factor $Q_{0}$ for all results.
	
	The form factor is defined in Eq. (\ref{eq_c_factor}).
	\begin{equation} \label{eq_c_factor}
		C = \frac{\abs{\int_{V} \dd^{3}x \mathbf{E}\cdot\mathbf{B} }^{2}}{\int_{V} \dd^{3}x \abs{\mathbf{B}}^{2}\int_{V} \dd^{3}x \varepsilon_{r} \abs{\mathbf{E}}^{2} } =  \frac{\abs{\int_{V} \dd^{3}x E_{z} }^{2}}{V\int_{V} \dd^{3}x \varepsilon_{r} \abs{\mathbf{E}}^{2} }
	\end{equation}
	The form factor is dependent on the distribution of the electric field vector $\mathbf{E}$ inside the cavity and therefore its value is different for each resonant mode of the cavity for a given external magnetic field distribution $\mathbf{B}$. The volume $V$ is defined to be the inside of the cavity and excludes the volume of conductors, such as a metal tuning rod. $\varepsilon_{r}$ is the relative permittivity which becomes relevant when there is dielectric material inside the cavity. The second equality in the equation is for when the magnetic field is constant and in one direction (denoted as the $z$-direction). Only the $E_{z}$ values of the electric field are important in this case.

	The dependence of $C$ and $Q_{0}$ for $R$ is shown below in Eq. (\ref{eq_scan_rate}).
	
	\begin{equation} \label{eq_scan_rate}
		R \propto Q_{0}C^{2}
	\end{equation}
	For most haloscopes with cylindrical cavities, $\mathbf{B}$ is relatively constant throughout the cavity volume and points in the axial direction of the cavity, allowing for the TM${}_{010}$ mode to be utilized for high $R$. In this setup, a cavity will have its frequency tuned by shifting the location of the tuning rod, following the mode which resembles the  TM${}_{010}$ mode throughout a range of frequencies.
	
	Ideally, both $C$ and $Q_{0}$ do not change much such that the whole frequency range with a tuning rod can probe an axion signal at the same sensitivity for a given amount of data integration time. In reality, however, this is not the case. When one uses a tuning rod, there needs to be a gap between the tuning rod and cavity end caps in order to ensure its movement, i.e., the length of the tuning rod needs to be slightly shorter than the height of the cavity. This means that these ``rod gaps'' between a metal tuning rod and cavity end caps may introduce unwanted capactive effects \cite{ADMX-Lyapustin} which perturb the electric field inside a cavity. This leads to a problem that is known as mode mixing. Orthogonal modes at the same resonance frequency are no longer independent of each other, and an unwanted mode will mix with the target resonant mode. This hybridization of modes causes a drastic decrease in both $C$ and $Q_{0}$, and not only is the scan rate suboptimal around this mode mixing region but there can also be a gap in frequencies the mode cannot reach at all. Most haloscope experiments have these mode mixing regions that are not sensitive to axions at the sensitivities of its surrounding frequencies. When haloscope experiments show exclusion results for the axion, mode mixings are responsible for leaving gaps in the exclusion plot. For example, in Refs. \cite{12TB-PRL, 12TB-PRD} there is a gap of frequencies around 1.094 GHz that did not have exclusion results due to mode mixings.
	
	In this work, we present a simple method to bypass the mode mixings for a cylindrical cavity with a copper tuning rod. This can be easily realized experimentally by inserting a dielectric rod inside the cavity. Then the cavity will become asymmetric and this makes a difference in the electric field when the tuning rod is moved in the direction closer to the dielectric insert compared to when it is farther away, even if the resonance frequency is identical. This changes the frequency at which the mode mixings occur, therefore making it possible to configure the cavity to scan through the full range of frequencies without losing sensitivity. We will also show that this configuration works even for higher frequencies when the dimensions of the cavity components are scaled down while keeping certain dimensions realistic.
	
	\begin{figure}[h]
		\centering
		\includegraphics[width=0.45\columnwidth]{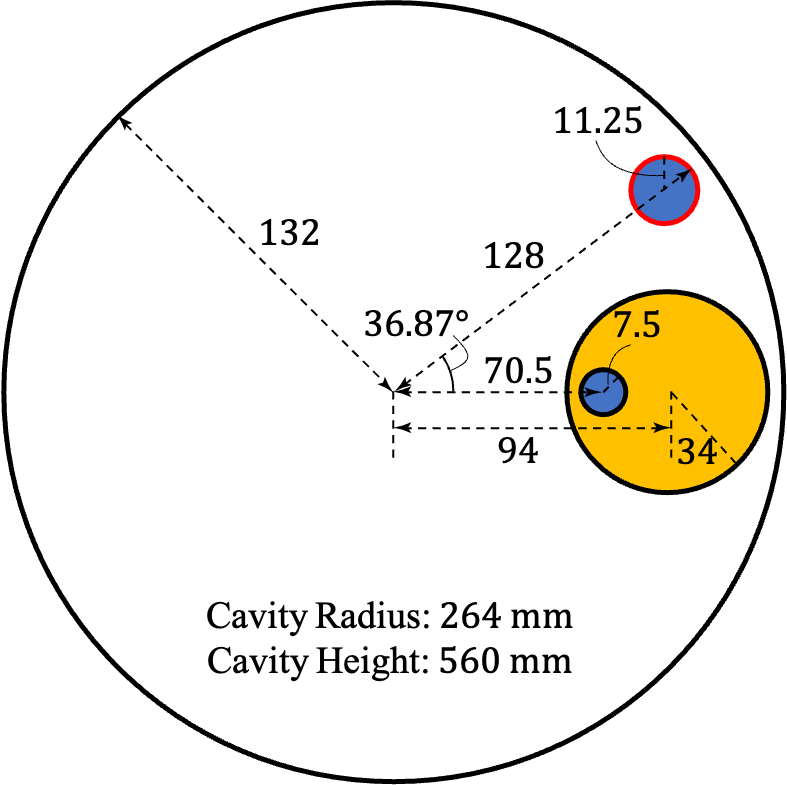} \quad
		\includegraphics[width=0.45\columnwidth]{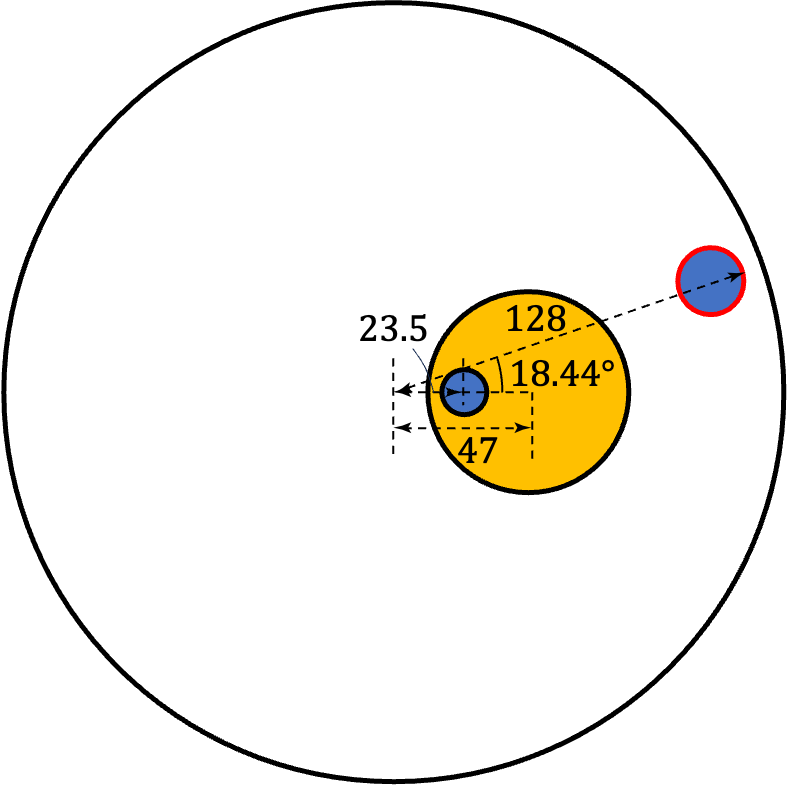}
		\caption
		{Design of the cavity with an dielectric insert. For the lower frequency range (left), the copper tuning rod (yellow) is located further away from the center of the cavity. The tuning axle is shown as a blue circle off-axis to the tuning rod. A dielectric insert (blue with red edges) is located where the unwanted modes appear in mode mixings. For the higher frequency range (right), the copper tuning rod is moved closer to the center of the cavity, and the dielectric insert is moved to a smaller angle. All dimensions are in mm.}
		\label{fig_design}
	\end{figure}
	
	\section{Asymmetric design for a cylindrical cavity}
	To begin with, we focus on a cylindrical cavity with an inner diameter of 132 mm and a tuning rod with an outer diameter of 34 mm, as seen in Fig. \ref{fig_design}. The height of the cavity is 560 mm and the tuning rod is 559 mm lengthwise. Due to the differences in the dimensions, there is a 0.5 mm gap between the end caps of the cavity and tuning rod. The tuning rod rotates off-axis with an axle which is located 70.5 mm from the center of the cavity, and 23.5 mm from the center of tuning rod. The resonance frequency of the cavity mode in this configuration ranges from 0.98 to 1.18 GHz. The cavity also has a higher frequency range when the axle is moved 47 mm closer to the center of the cavity, and the frequency range in this case is 1.18 -- 1.44 GHz. With the exception of the inner cavity being 2 mm larger for this work, these are the same dimensions as in Ref. \cite{12TB-Tuning}, and its capability of physically rotating the tuning rod has also been shown.
	
	This cavity design also includes a dielectric insert. This insert is desgined to be fixed at a set location that does not interfere with the movement of the tuning rod. The height of the dielectric insert is the same as the cavity, without any gaps, and the diameter is 11.25 mm. The insert is 4 mm away from the cavity walls at its closest point, but at an 36.87$\degree$ angle in comparison to the tuning rod's location when it is closest to the edge of the cavity. For the higher frequency configuration, the dielectric insert is moved to an angle of 18.44$\degree$ although the distance from the center of the cavity is the same. Both configurations with the dielectric insert can be seen in Fig. \ref{fig_design}.
	
	By using finite-element simulation \cite{CST}, the electric field distribution of the cavity mode is obtained along with the quality factor of the cavity assuming that its walls and the tuning rod are made out of copper and measured at room temperature. The dielectric insert's properties are modeled to follow the properties of alumina, which has a dielectric constant of 9.9. Simulations were done for both the lower and higher frequency configurations and the TM${}_{010}$-like mode of the cylindrical cavity was analyzed. Such modes were found by looking for the resonant mode with the highest form factor $C$ for each location of the tuning rod, even when there were mode mixings. The form factor was calculated using Eq. (\ref{eq_c_factor}). The electric field was obtained by exporting the electric field vector values into a grid of cells which are 4 mm wide in all directions. Using 4 mm instead of 2 mm greatly reduces the computation time when calculating $C$ as it scales with the number of voxels, which is inversely proportional to the cube of the grid size. For this reason, the cavity radius in this work was increased to 264 mm, a value divisible by four. The magnetic field was assumed to be a constant value in the $z$-direction.
	
	\begin{figure}[h]
		\centering
		\includegraphics[width=0.8\columnwidth]{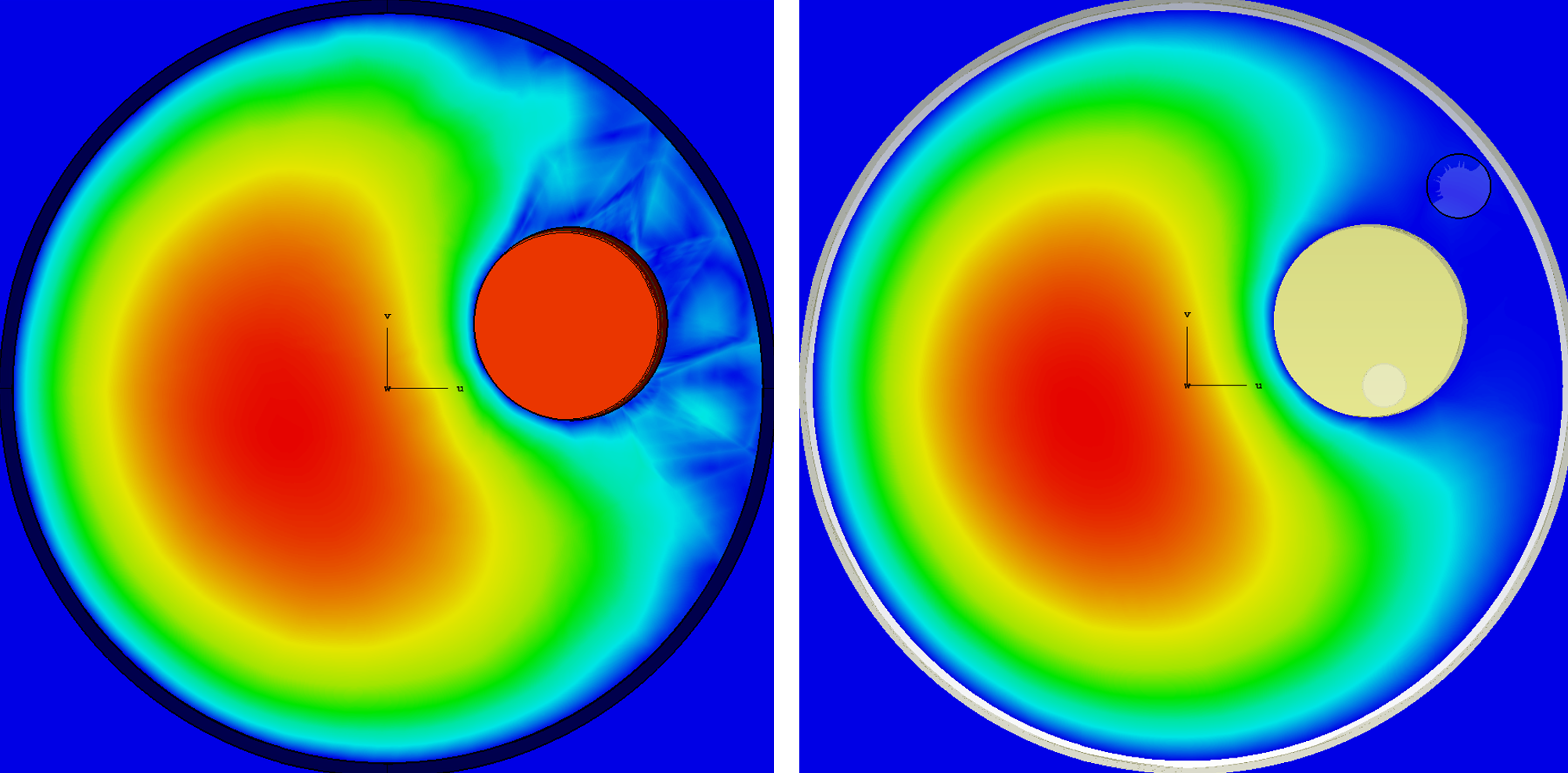}
		\caption
		{The left figure shows the electric field distribution when there is a mode mixing (seen in the top right side of the cavity) without the dielectric insert. The right figure shows the simulation results at the same location but with the dielectric insert (top right circle with black edges). The mode mixing no longer exists. The $u$, $v$, and $w$ axes shown in the figures correspond to $x$, $y$, and $z$ in Cartesian coordinates. Both results were obtained through finite-element simulation.}
	\label{fig_mode_mixings}
	\end{figure}
	
	The significance in this simplistic design change lies in the fact that the cavity is no longer symmetrical. Also, the dielectric is strategically located at where the hybrid mode appears when there is a mode crossing. By doing this, the insert will disrupt the unwanted hybrid mode that would have been localized in its absence, as illustrated in Fig. \ref{fig_mode_mixings}. The change in electric field also changes the frequency of the mode mixing. Comparing this to the case where the tuning rod is on the opposite side (i.e., the same rotation angle in the opposite direction), the mode mixing frequency does not change as much since the electric field is weaker where the dielectric insert is located.
	
	\begin{figure}[h]
		\centering
		\includegraphics[width=\columnwidth]{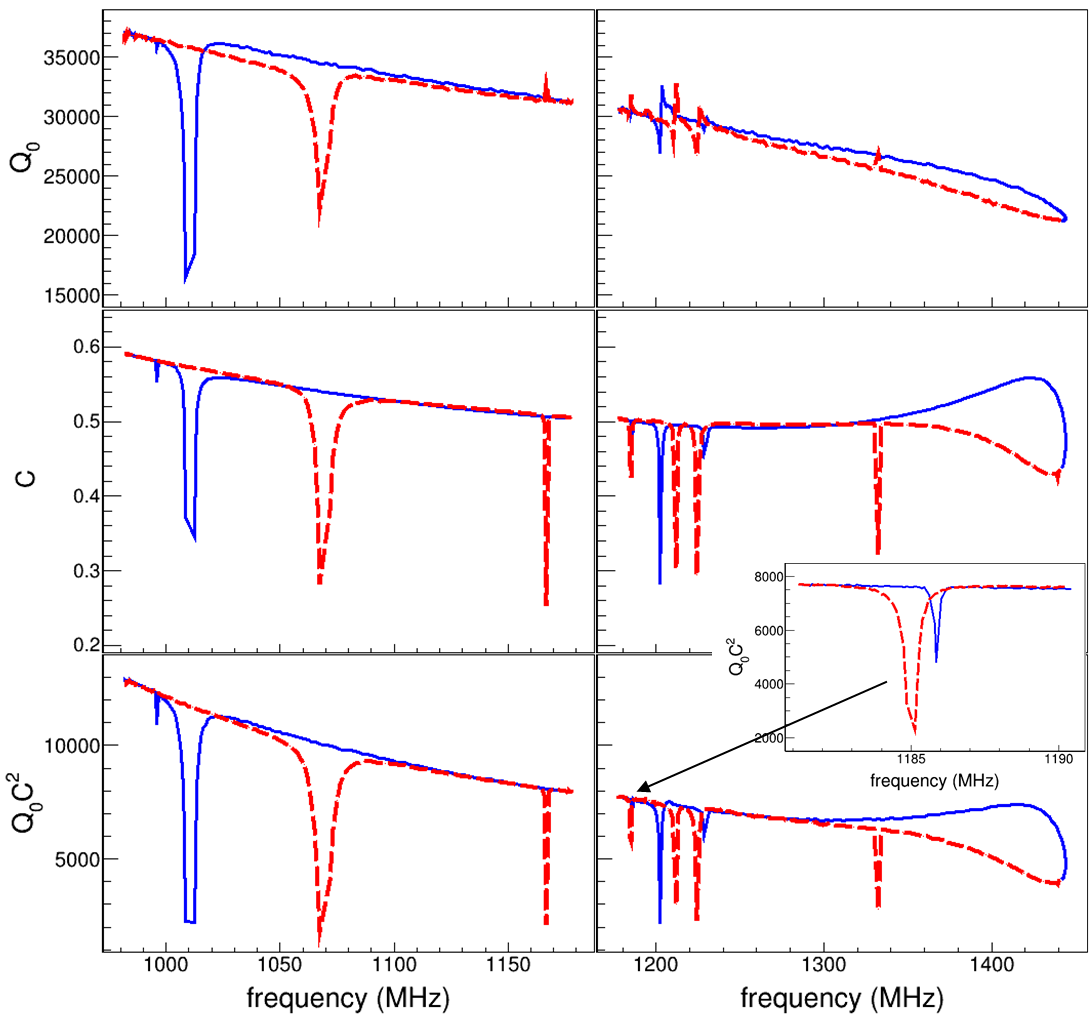}
		\caption
		{Simulation results for the asymmetrical cavity design with an inner diameter of 264 mm. The top, middle, and bottom figures show the unloaded quality factor $Q_{0}$, form factor $C$, and a proportional measure of the figure of merit $Q_{0}C^{2}$, respectively. The left and right side of each figure correspond to the lower and higher frequency range configurations. The inset figure at the middle right shows a close-up for $Q_{0}C^{2}$ around 1.185 GHz, which is where the arrow points to in the wider range, with finer resolution between rotation angles. For all figures, the red dashed line is when the tuning rod moves away from the dielectric insert, while the blue line is when the tuning rod moves closer to it.}
		\label{fig_264mm}
	\end{figure}
	
	The simulation-obtained unloaded quality factor $Q_{0}$ and form factor $C$, along with the value $Q_{0}C^{2} \propto R$, are shown in Fig. \ref{fig_264mm} for both the lower and higher frequency ranges. The narrow dips in these parameters signify a mode mixing. In these regions the converted axion signal power will be much lower and therefore it is not viable for scanning. However, by selectively alternating between the rotation direction of the tuning rod which has a better $Q_{0}C^{2}$, it is possible to retain a relatively steady scan rate that is unhindered by the loss from mode mixings. Therefore, one can conduct an axion search experiment for a wide range of frequencies without any gaps and the only change in cavity configuration is at 1.18 GHz when both the tuning rod and dielectric insert change positions. Moreover, since this design also reuses the same tuning rod and dielectric insert for both configurations, it is not necessary to manufacture multiple tuning rods or related parts to run the experiment for an extended frequency range.
	
	The idea of using an asymmetric configuration inside a cavity has been employed in previous axion dark matter searches such as Refs. \cite{HAYSTAC-Inst} and \cite{ADMX-DFSZ2}. In particular, the former work uses a tunable dielectric rod placed in an asymmetric position compared to the copper tuning rod and shows the improvements of removing mode mixings when comparing between Refs. \cite{HAYSTAC} and \cite{HAYSTAC-P1}. Both experiments have made use of their cavity designs to enhance frequency coverage. 
	
	While this work is similar in terms of design, the dielectric insert is a static object inside the cavity strategically placed to circumvent almost all of the negative effects stemming from mode mixings. Therefore this configuration provides an alternative that removes the need of requiring to mechanically operate a second motorized object inside the cavity, which can potentially cause heat generation or frequency drifts depending on the stability of the tuning mechanism. Moreover, this is consistently possible throughout a frequency range of about 38\% with respect to the central frequency. This allows for wider coverage in parameter space using a single set of cavity, tuning rod, and dielectric insert with only one placement change of components in the middle of the frequency range.
	
	Using this design can help to conduct a continuous axion haloscope experiment throughout a cavity's full frequency range without having to sacrifice sensitivity at certain mode mixing frequencies.
	
	\section{Extension to higher frequency ranges}
	As seen in the previous section we have shown that it is possible to avoid mode mixings through an asymmetric design of the cavity. In this section we present the results of simulations where the cavities have been scaled down, which changes the frequency range.
	
		\begin{figure}[h]
		\centering
		\includegraphics[width=\columnwidth]{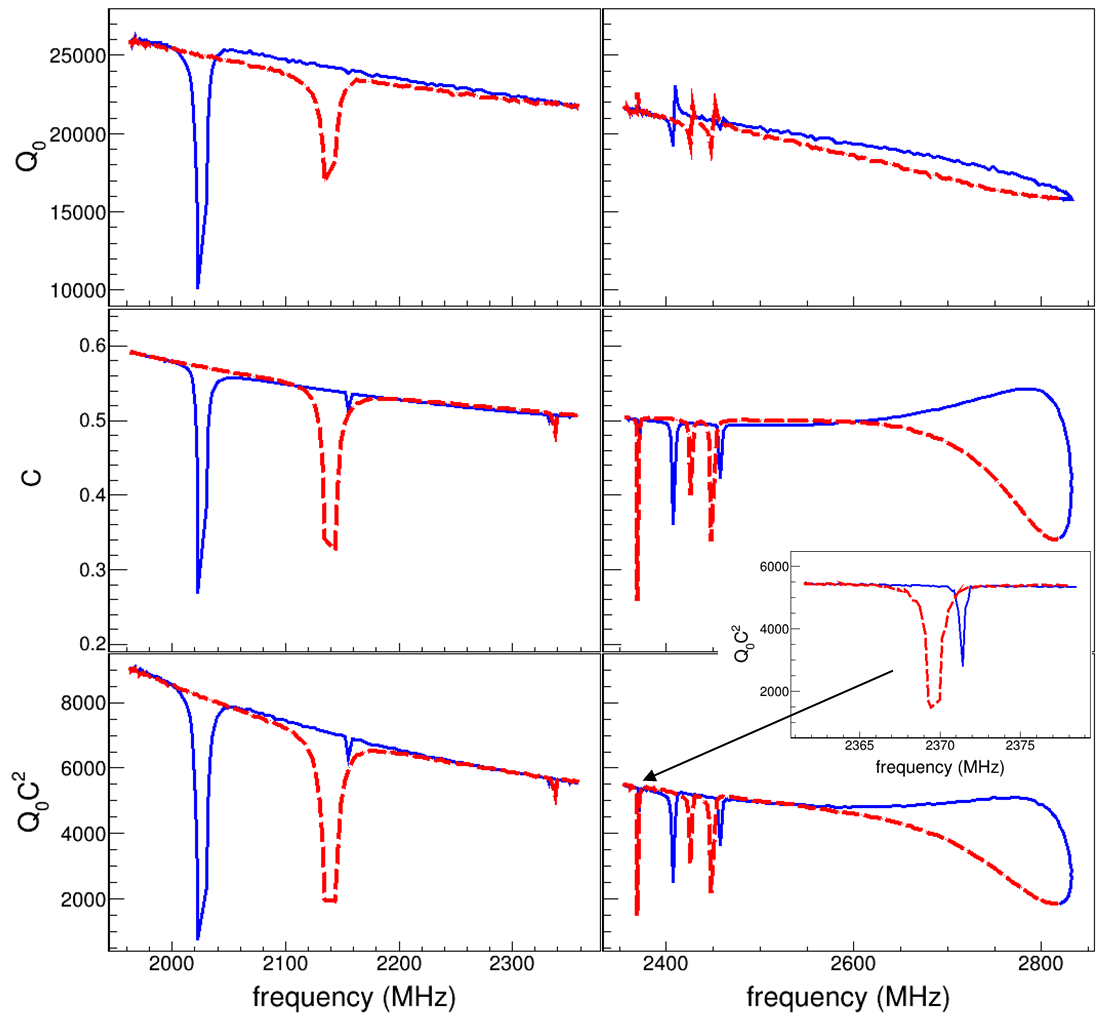}
		\caption
		{Simulation results for the asymmetrical cavity design with an inner diameter of 132 mm. The top, middle, and bottom figures show the unloaded quality factor $Q_{0}$, form factor $C$, and a proportional measure of the figure of merit $Q_{0}C^{2}$, respectively. The left and right side of each figure correspond to the lower and higher frequency range configurations. The inset figure at the middle right shows a close-up for $Q_{0}C^{2}$ around 2.37 GHz, which is where the arrow points to in the wider range, with finer resolution between rotation angles. For all figures, the red dashed line is when the tuning rod moves away from the dielectric insert, while the blue line is when the tuning rod moves closer to it.}
		\label{fig_132mm}
	\end{figure}
	
	\begin{figure}[h]
		\centering
		\includegraphics[width=\columnwidth]{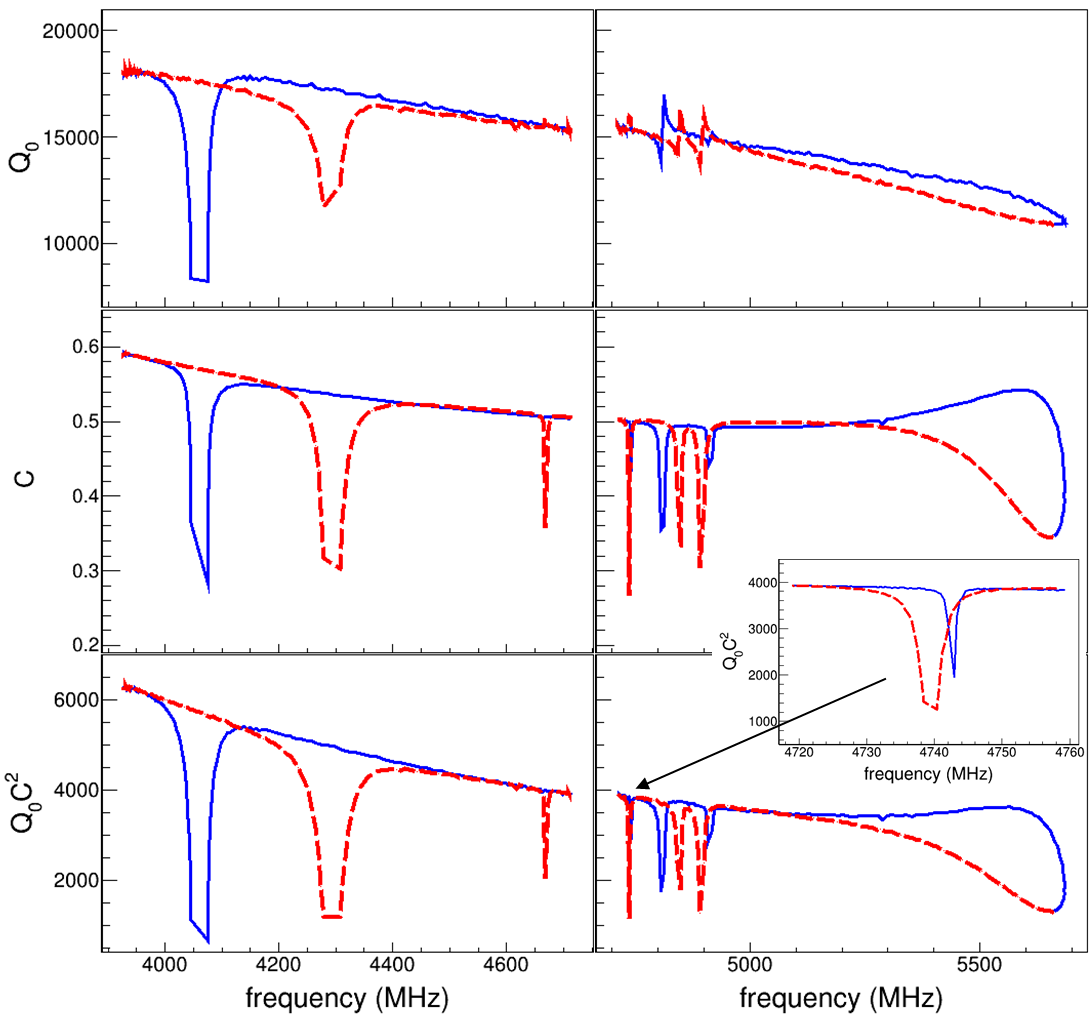}
		\caption
		{Simulation results for the asymmetrical cavity design with an inner diameter of 66 mm. The gap between the tuning rod and end caps are not scaled and kept at 0.25 mm. The top, middle, and bottom figures show the unloaded quality factor $Q_{0}$, form factor $C$, and a proportional measure of the figure of merit $Q_{0}C^{2}$, respectively. The left and right side of each figure correspond to the lower and higher frequency range configurations. The inset figure at the middle right shows a close-up for $Q_{0}C^{2}$ around 4.74 GHz, which is where the arrow points to in the wider range, with finer resolution between rotation angles. For all figures, the red dashed line is when the tuning rod moves away from the dielectric insert, while the blue line is when the tuning rod moves closer to it.}
		\label{fig_66mm}
	\end{figure}
	
	Shown in Fig. \ref{fig_132mm} are the results of a cavity that is reduced to half the size, making the inner diameter of the cavity 132 mm. This requires the gap between the tuning rod and the cavity end caps to be 0.25 mm. This is similar to a 0.01 inch (0.254 mm) gap, which has been shown to be possible in other experiments such as in Ref. \cite{HAYSTAC-BeadPull}. The frequency ranges for the cavity are 1.96 -- 2.36 GHz and 2.36 -- 2.82 GHz respectively, depending on the tuning rod axle's location. The asymmetric design shows that even for this smaller cavity the locations of mode mixings are at different frequencies based on whether the rod is moved closer to or further away from the dielectric insert. Calculations for the form factor use a grid size of 2 mm for the electric field, assuming the constant magnetic field.
	
	In Fig. \ref{fig_66mm}, we continue to reduce the size of the cavity, now scaling down all dimensions by two once more. The inner diameter of the cavity is 66 mm. There is one exception to the scaling, which is the rod gap. This will be maintained at 0.25 mm, reflecting any precision issues that can occur in manufacturing due to a small rod gap. Despite the gap being larger relative to other dimensions, the mode mixings are still moved to different frequencies depending on the direction of rotation. Calculations for the form factor use a grid size of 1 mm for the electric field, assuming the constant magnetic field.
	
	\section{Conclusion}
	When tuning a cavity, there inevitably are some portions of the resonance frequency range where the mixing of modes significantly degrades the sensitivity to the axion. This is due to the decrease in form factor and quality factor of the cavity, among other parameters. In order to bypass these issues, we have created and simulated an asymmetric cavity design which places a dielectric insert that shifts the location of the mode mixing frequencies. This simple geometry of this design can be easily implemented in axion haloscope experiments and allows one to choose between the tuning rod rotation directions for the same resonance frequency in order to have a consistently high scan rate across the full frequency range. In addition, the asymmetrical design can be replicated for cavities of higher frequencies by scaling down the cavity size, even if the gap between the tuning rod and cavity endcaps cannot be reduced to the same scale and is relatively larger. Therefore, the design presented in this work is not limited to a single size and can be adapted to various cylindrical cavity diameters that can be manufactured.
	
	\section*{Acknowledgements}
	This work was supported by a Korea University Grant and NRF-1234567890.


\begin{thebibliography}{99}
	\bibitem{Axion1}
	S. Weinberg, A New Light Boson?, Phys. Rev. Lett. 40 (1978) 223.
	
	\bibitem{Axion2}
	F. Wilczek, Problem of Strong P and T Invariance in the Presence of Instantons, Phys. Rev. Lett. 40 (1978) 279.
	
	\bibitem{StrongCP1}
	G. 't Hooft, Symmetry Breaking Through Bell-Jackiw Anomalies, Phys. Rev. Lett, 37 (1976) 8.
	
	\bibitem{StrongCP2}
	G. 't Hooft, Computation of the Quantum Effects Due to a Four-Dimensional Pseudoparticle, Phys. Rev. D 14 (1976) 3432 [Erratum ibid. 18 (1978) 2199].
	
	\bibitem{StrongCP3}
	J.H. Smith, E.M. Purcell, and N.F. Ramsey, Experimental limit to the electric dipole moment of the neutron, Phys. Rev. 108 (1957) 120.
	
	\bibitem{StrongCP4}
	W.B. Dress, P.D. Miller, J.M. Pendlebury, P. Perrin, and N.F. Ramsey, Search for an Electric Dipole Moment of the Neutron, Phys. Rev. D 15 (1977) 9.
	
	\bibitem{StrongCP5}
	I.S. Altarev et al., A search for the electric dipole moment of the neutron using ultracold neutrons, Nucl. Phys. A341 (1980) 269. 
	
	\bibitem{PQ}
	R.D. Peccei and H.R. Quinn, CP Conservation in the Presence of Instantons, Phys. Rev. Lett. 38 (1977) 1440.
	
	\bibitem{AxionCDM1}
	J. Preskill, M.B. Wise, and F. Wilczek, Cosmology of the Invisible Axion, Phys. Lett. 120B (1983) 127
	
	\bibitem{AxionCDM2}
	L.F. Abbott and P. Sikiviem, A Cosmological Bound on the Invisible Axion,  Phys. Lett. B120 (1983) 133
	
	\bibitem{AxionCDM3}
	M. Dine and W. Fischler, The Not So Harmless Axion, Phys. Lett. B120 (1983) 137
	
	\bibitem{CDM1}
	V. Rubin and W.K. Ford Jr., Rotation of the Andromeda Nebula from a Spectroscopic Survey of Emission Regions, ApJ 159 (1970) 379.
	
	\bibitem{CDM2}
	Douglas Clowe et al., A Direct Empirical Proof of the Existence of Dark Matter, ApJ 648 (2006) L109.
	
	\bibitem{Planck}
	Planck collaboration, P.A.R. Ade et al., Planck 2015 results. XIII. Cosmological parameters, Astron. Astrophys. 594 (2016) A13. 
	
	\bibitem{Sikivie}
	P. Sikivie, Experimental Tests of the Invisible Axion, Phys. Rev. Lett. 51 (1983) 1415; Phys. Rev. D 32 (1985) 2988.
	
	\bibitem{Cavity}
	C. Hagmann,  P. Sikivie,  N. Sullivan,  D. B. Tanner, and  S.-I. Cho, Cavity design for a cosmic axion detector, Rev. Sci. Instrum. 61 (1990) 1076.
	
	\bibitem{KSVZ1}
	J.E. Kim, Weak-Interaction Singlet and Strong CP Invariance, Phys. Rev. Lett. 43 (1979) 103.
	
	\bibitem{KSVZ2}
	M.A. Shifman, A.I. Vainshtein, and V.I. Zakharov, Can confinement ensure natural CP invariance of strong interactions? Nucl. Phys. B166 (1980) 493.
	
	\bibitem{DFSZ1}
	A.R. Zhitnitskii, On Possible Suppression of the Axion Hadron Interactions, Yad. Fiz. 31 (1980) 497 [Sov. J. Nucl. Phys. 31 (1980) 260].
	
	\bibitem{DFSZ2}
	M. Dine, W. Fischler, and M. Srednicki, A simple solution to the strong CP problem with a harmless axion, Phys. Lett. 104B (1981) 199.
	
	\bibitem{ScanRate}
	L. Krauss, J. Moody, F. Wilczek and D.E. Morris, Calculations for cosmic axion detection, Phys. Rev. Lett. 55 (1985) 1797.
	
	\bibitem{ADMX-Lyapustin}
	D. Lyapustin, An improved low-temperature RF-cavity search for dark-matter axions, Doctoral dissertation, University of Washington, (2015).
	
	\bibitem{12TB-PRL}
	A.K. Yi et al., Axion Dark Matter Search around 4.55 $\mu$eV with Dine-Fischler-Srednicki-Zhitnitskii Sensitivity, Phys. Rev. Lett. 130 (2023) 071002.
	
	\bibitem{12TB-PRD}
	A.K. Yi et al., Search for the Sagittarius tidal stream of axion dark matter around 4.55 $\mu$eV, Phys. Rev. D 108 (2023) L021304.
	
	\bibitem{12TB-Tuning}
	A.K. Yi et al., Enhanced tunable cavity development for axion dark matter searches using a piezoelectric motor in combination with gears, JINST 19 (2024) T07004.
	
	\bibitem{CST}
	www.cst.com.
	
	\bibitem{HAYSTAC-Inst}
	S. Al Kenany et al., Design and operational experience of a microwave cavity axion detector for the 20--100 $\mu$eV range, Nucl. Instrum. Methods Phys. Res., Sect. A 854 (2017) 11.	
	
	\bibitem{ADMX-DFSZ2}
	ADMX collaboration, T. Braine et al., Extended Search for the Invisible Axion with the Axion Dark Matter Experiment, Phys. Rev. Lett. 124 (2020) 101303.
	
	\bibitem{HAYSTAC}
	B. M. Brubaker et al., First Results from a Microwave Cavity Axion Search at 24 $\mu$eV, Phys. Rev. Lett. 118 (2017) 061302.
	
	\bibitem{HAYSTAC-P1}
	L. Zhong et al., Results from phase 1 of the HAYSTAC microwave cavity axion experiment, Phys. Rev. D 97 (2018) 092001.
	
	\bibitem{HAYSTAC-BeadPull}
	N.M. Rapidis, S.M. Lewis and K.A. van Bibber, Characterization of the HAYSTAC axion dark matter search cavity using microwave measurement and simulation techniques, Rev. Sci. Instrum. 90 (2019) 024706.
	
\end{thebibliography}
\end{document}